\documentclass[twocolumn]{aastex63}

\usepackage{gensymb}
\usepackage[euler]{textgreek}

\usepackage{color,soul}
\definecolor{red}{rgb}{0.5,0,0}
\setulcolor{red}

\usepackage{xcolor}

\accepted{September 15, 2021}
\submitjournal{ApJL}

\shorttitle{Signatures on Detached Kuiper Belt}
\shortauthors{Anderson \& Kaib}

\graphicspath{{./}{figures/}}

\begin{document}

\title{Signatures of a Distant Planet on the Inclination Distribution of the Detached Kuiper Belt}

%\correspondingauthor{Kalee Anderson}
%\email{kaleeanderson@ou.edu}

\author{Kalee E. Anderson}
\affiliation{HL Dodge Department of Physics and Astronomy \\ University of Oklahoma \\ Norman, OK 73019, USA}

\author{Nathan A. Kaib}
\affiliation{HL Dodge Department of Physics and Astronomy \\ University of Oklahoma \\ Norman, OK 73019, USA}

\begin{abstract}

A distant, massive planet in the outer solar system has recently been proposed to explain some observed features of extreme trans-Neptunian objects (TNOs). Here we use N-body simulations of the formation of the Kuiper belt and Oort cloud as well as a survey simulator to compare models of the solar system with and without a 9th planet to one another as well as to observations. The main mechanism for TNOs to be deposited into the distant ($a$\textgreater 50 au), detached ($q$ \textgreater 40 au) region of the Kuiper Belt in the 8-planet model is Kozai-Lidov oscillation of objects in mean motion resonances (MMR) with Neptune. This effect does not deposit low-inclination ($i \lesssim$ 20\degree) objects into this region. However, we find that the 9th planet generates a group of distant, detached TNOs at low inclinations that are not present in the 8-planet model. This disparity between the 8-planet and the 9-planet models could provide a strong constraint on a possible planet 9 with further detections of TNOs in the distant, detached region of the Kuiper Belt. 

\end{abstract}

\section{Introduction} \label{sec:intro}

Recently a distant, 5-10 $M$\textsubscript{\(\oplus\)} planet has been proposed to explain some puzzling features of the known population of trans-Neptunian objects (TNOs). The planet provided an explanation for an observed asymmetry in the orbital distribution of TNOs with large (with respect to the classical belt) semimajor axis ($a$ $\gtrsim$ 250 au) that have perihelia ($q$) large enough to be dynamically detached from the known planets ($q \gtrsim$ 40 au) \citep{2014Natur.507..471T,2016AJ....151...22B}.  A distant ninth planet could also explain the observed population of high-inclination ($i \gtrsim$ 45\degree) scattering objects, as scattering interactions with giant planets are not very effective at raising the inclinations of TNOs \citep{2012MNRAS.420.3396B}, and perturbations from a distant massive planet would enhance production of high-inclination scattering objects \citep{2016ApJ...833L...3B,2019AJ....158...43K}. The distant planet has been proposed to have a mass of $\sim$ 5-10 $M$\textsubscript{\(\oplus\)}, semimajor axis of $a$ $\sim$ 500-1000 au, eccentricity of $e \sim$ 0.2-0.7, and inclination of $i \sim$ 20\degree  \citep{2016AJ....151...22B,2019PhR...805....1B}.

The inclination distributions of detached and scattering bodies have been previously explored as a constraint on the properties of a possible distant planet, such as the plane of the planet \citep{2016ApJ...824L..23B}. However, the inclinations of distant, detached TNOs may be a particularly powerful tool that can be developed to further constrain the properties of this possible planet. During Neptune’s migration, TNOs can be captured into distant Neptunian resonances \citep{2015AJ....149..202P,2016AJ....152...23V}. If such TNOs reach a critical inclination, they can begin to undergo Kozai-Lidov cycles \citep{1962AJ.....67..591K,1962P&SS....9..719L,2003EM&P...92...29G}. During these cycles the object's eccentricity and inclination oscillate exactly out of phase. At high inclination, the perihelion increases (while semimajor axis remains fixed), weakening the resonant coupling with Neptune. During this phase, the object can exit the Neptunian resonance and get locked in its orbit, a marking of the former location of Neptune's MMR as the actual MMRs continue to migrate with Neptune. This process can produce a high-inclination dynamically inert population near modern MMRs with stable high perihelion orbits, detached from the planets \citep{2016AJ....152..133K,2016ApJ...827L..35N}. This inactive high perihelion population will also have high inclination due to the past Kozai cycles having raised the inclination as the perihelion increased, with 99\% of $q$ \textgreater 40 au objects predicted to have inclinations greater than 20\degree \citep{2016AJ....152..133K}. 

Because the Kozai mechanism only produces high-inclination orbits when populating the distant detached region, the low-inclination region should not be populated as is seen in the 8-planet model in Figure \ref{fig:scatter} denoted by the red box. However, a distant ninth planet would perturb the high perihelion population as well. Secular perturbations from a distant planet drive changes in both the perihelia and inclinations of large ($\gtrsim$100 AU) semimajor axis TNOs \citep{2016AJ....151...22B,2017AJ....154..229B,2019PhR...805....1B}. These secular interactions are not limited to high-inclination orbits, so the low-inclination region left empty from Kozai cycles in an 8-planet system could be populated due to influences of a distant ninth planet, as can be seen by the particles that populate the region denoted by the red boxes in the 9-planet models in Figure \ref{fig:scatter}. The inclination distribution of detached TNOs with large ($\gtrsim$ 100 AU) semimajor axes could, thus, provide a powerful new constraint on a ninth planet's properties as well as its existence. Only a modest number of presently known TNOs have $q$ \textgreater 40 AU and $a$ $\gtrsim$100 AU, but the Vera C. Rubin Observatory is poised to detect large numbers of such bodies in the very near future, providing powerful new constraints on any distant undiscovered planets \citep{2019ApJ...873..111I}.

In this work, we use N-body simulations of the formation of the Kuiper belt and Oort Cloud with observational bias taken into account to compare to detected TNOs. We run simualtions with and without a ninth planet to develop new constraints on a possible distant, undiscovered planet. This work is organized into the following sections. In Section \ref{sec:method}, we describe our numerical simulations and simulated detections. In Section \ref{sec:results}, we present our results. In Section \ref{sec:con}, we discuss the implications of our work.

\section{Method} \label{sec:method}

\subsection{Numerical Simulations\label{subsec:simulations}}

We perform numerical simulations of the formation of the Kuiper belt and Oort Cloud using the SWIFT RMVS4 integrator \citep{1994Icar..108...18L} using four different outer-planet configurations. Each simulation has Jupiter, Saturn, and Uranus starting with their current semi-major axes and small eccentricities ($e$ \textless  0.01) and inclinations ($i$ \textless  1 \degree). Neptune starts at 24 au with a small eccentricity and inclination. Three of the simulations also include a proposed fifth giant planet, each with a different orbit. Each simulation also has a disk of 1 million test particles. The implantation efficiency of the test particles into the different parts of the Kuiper Belt is $\sim 10^{-3}$ to $10^{-4}$, which necessitates such a large number \citep{2016ApJ...825...94N}. The test particles are randomly drawn from a uniform semimajor axis distribution of 24-30 au (corresponding to a $a^{-1}$ surface density profile) and a uniform eccentricity distribution between 0 and 0.01. The inclinations of the test particles are randomly drawn from the function 
\[ f(i) = \sin{i} \exp - \frac{i}{2 \sigma^2 }  \] 
with $\sigma = 1 \degree$. The other orbital elements are randomly drawn from an isotropic distribution. 

In each simulation, Neptune migrates through this disk of particles similar to the late proposals of the Nice model \citep{2012AJ....144..117N}. We give Neptune the same migration pattern employed in \citet{2016AJ....152..133K} which was motivated by the works of \citet{2015AJ....150...73N,2015AJ....150...68N} and \citet{2016ApJ...825...94N}. Neptune migrates from 24 to 28 au. It then jumps by 0.5 au in semimajor axis and to an eccentricity of 0.05. This is then followed by a final migration from 28.5 AU to Neptune's current orbit. Neptune's jump in semimajor axis and eccentricity is meant to mimic a gravitational scattering event with an ejected planet \citep{2011ApJ...742L..22N,2015AJ....150...68N,2012AJ....144..117N}. Prior to the jump, Neptune migrates with an e-folding timescale of 30 Myr and afterward a post-jump migration timescale of 100 Myr \citep{2016AJ....152..133K}. After Neptune has reached its modern semi-major axis, migration stops and the simulation is run up to a total simulation time of 4 Gyr with a 200 day timestep. 

As in the ``grainy slow simulation" in \citet{2016AJ....152..133K}, Neptune's migration was ``grainy,'' with thousands of small ($\delta a \sim 10^{-3}$ au) sudden shifts in semi-axis to simulate Neptune having close encounters with $\sim$ 2000 Pluto-mass objects proposed to have been present in the primordial Kuiper Belt \citep{2016ApJ...825...94N}. Perturbations from passing field stars \citep{2008CeMDA.102..111R} and the Galactic tide \citep{1986Icar...65...13H} were also included in the simulations. These external perturbations are known to detach the perihelia of distant solar system bodies and are therefore important to include in our simulations \citep{2011Icar..215..491K}.

We compare the 4-giant-planet simulation with the 3 simulations that include an additional distant planet. In the high-eccentricity P9 simulation, the distant planet has an eccentricity of 0.6, an inclination of 20\degree , a mass of 10 $M\textsubscript{\(\oplus\)}$, and a semimajor axis of 700 au \citep{2018AJ....155..250K}. In the low-eccentricity P9 simulation, the distant planet has an eccentricity of 0.25, an inclination of 20\degree , a mass of 5 $M$\textsubscript{\(\oplus\)}, and a semimajor axis of 500 au \citep{2019PhR...805....1B}. In the circular P9 simulation we employ a conservative, ``least perturbative" case for the distant planet with a circluar orbit with an eccentricity of $10^{-3}$ and an inclination of \textless  1\degree , a mass of 5 $M$\textsubscript{\(\oplus\)}, and a semimajor axis of 700 au. Although such an orbit and mass have not been proposed, we take this to be a limiting ``least perturbative" undiscovered-planet model as the circular, low inclination, and distant orbit will, on average, generate weaker perturbation on TNOs than a closer, eccentric, highly inclined orbit.

\subsection{Survey Simulator\label{subsec:sursim}}

In order to directly compare our models to known TNOs, observational bias must be accounted for. We utilize the survey simulator developed by the Outer Solar System Origins Survey (OSSOS) and Canada-France Ecliptic Plane Survey (CFEPS) teams \citep{2018FrASS...5...14L}. This simulator applies the survey bias of the well-characterized OSSOS survey (and its predecessor surveys) to the input model. To do this, the simulator randomly draws an object from our numerical model and then assigns it an absolute magnitude in the $r$ band ($H\textsubscript{r}$) by randomly sampling from a given $H\textsubscript{r}$ distribution. Given the $H\textsubscript{r}$ and the object’s position (given by its orbit), the survey simulator determines if the object would be observed and tracked for three detections. 

We use two different $H\textsubscript{r}$ distributions that have previously been proposed. The first is a “contrast” distribution as described in \citet{2018AJ....155..197L}. This distribution is defined with a disjointed power law ( \( \frac{dN}{dH_{r}}  \propto 10^\alpha \) ). The bright end ($H\textsubscript{r}$ \textless  8.3) has a power law of \textalpha = 0.9. At $H\textsubscript{r}$ = 8.3 there is a discontinuous drop of a factor of 3 and then a faint-end power law of \textalpha = 0.4 ($H\textsubscript{r}$ \textgreater 8.3) is used for fainter objects. The second is a “knee” distribution \citep{2018AJ....155..197L}. This distribution is similar to the contrast without the discontinuous drop. The bright end ($H\textsubscript{r}$ \textless 7.7) has a power law of \textalpha = 0.9. Then at $H\textsubscript{r}$=7.7 there is a continuous transition to a faint-end power law of \textalpha = 0.4.

The survey simulator with $H\textsubscript{r}$ distributions shows how the numerical simulation populations would be distributed if observed by OSSOS and its predecessor surveys \citep{2011AJ....142..131P,2017AJ....153..236P,2016AJ....152..111A,2018ApJS..236...18B}. However, these surveys only detected a handful of the detached objects in which we are interested (we limit ourselves to these detections since they originate from such well-characterized surveys). As explained in the subsequent sections, small number statistics strongly limit the confidence with which we can confirm or invalidate our different numerical models. Nevertheless, in this work we highlight how this particular population of bodies is quite sensitive to the existence of an additional, undetected distant planet.

\section{Results} \label{sec:results}

\begin{deluxetable}{lccc}
\tablenum{1}
\tablecaption{OSSOS-detected Objects in Our Area of Interest \label{tab:OSSOSobj}}
\tablewidth{5pt}
\tablehead{
\colhead{Object} & \colhead{a} & \colhead{q} &
\colhead{i}
}
%\decimalcolnumbers
\startdata
2015 RB\textsubscript{278} & 75.66 & 42.45 & 27.70 \\
2015 GB\textsubscript{56} & 91.07 & 41.54 & 29.46 \\
2015 KE\textsubscript{172} & 129.80 & 44.13 & 38.36 \\
2013 GP\textsubscript{136} & 150.24 & 41.03 & 33.54 \\
2013 UT\textsubscript{15} & 200.26 & 43.93 & 10.65 \\
\enddata
\tablecomments{5 OSSOS-detected objects in our area of interest: semimajor axes between 75 and 250 AU and perihelion between 40 and 50 AU. Columns are (1) object name, (2) semimajor axis, (3) perihelion, and (4) inclination. }
\end{deluxetable}

There are 5 OSSOS-detected objects in our area of interest, shown in Table \ref{tab:OSSOSobj}. We use these five objects to compare with the resulting orbital distributions of our numerical simulations. Previous studies by \citet{2016AJ....152..133K} and \citet{2016ApJ...827L..35N} focused on the production of detached TNOs under the influence of the known giant planets. In their works, these objects only reside at high inclinations when their semimajor axes are near mean motion resonances with Neptune. In this work, we look at how a distant massive planet affects the inclination distribution of detached, distant Kuiper Belt objects. We define detached as $q$ \textgreater 40 au. To limit to objects that can be detected reasonably, we also use an upper cutoff of $q$ \textless 50 au. 

\begin{figure*}
\includegraphics[width=\textwidth]{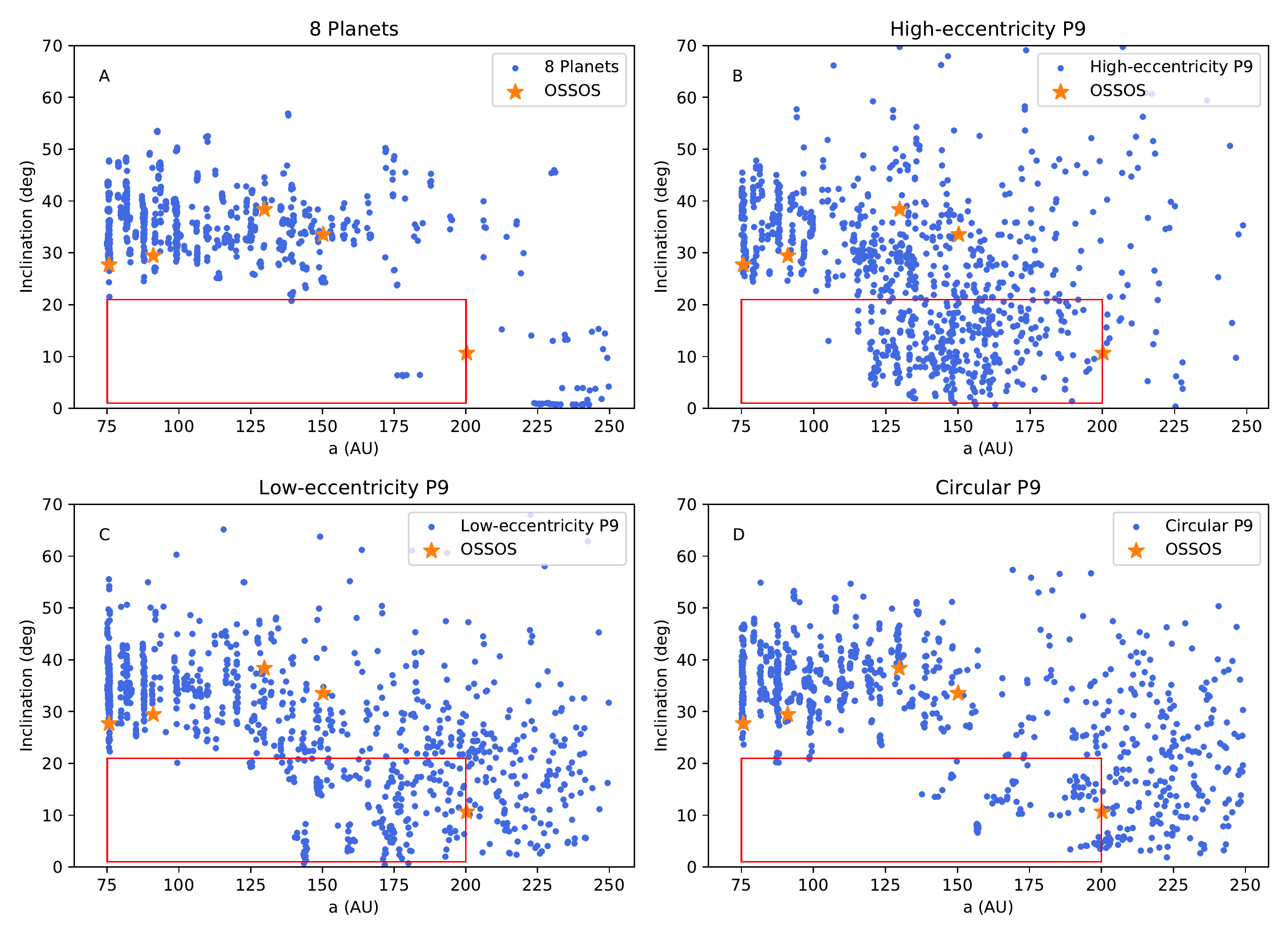}
%\includegraphics{scatter.pdf}
%%\plotone{scatter.pdf}
\caption{Plots of simulated detections generated from numerical simulation outputs run through the OSSOS survey simulator assuming the contrast size distribution. Simulated objects are blue and detected OSSOS objects are marked with orange stars. Orbital inclination is plotted vs. semimajor axis. Panel A is the 8 planet simulation, panel B is the high-eccentricity P9 simulation, panel C is the low-eccentricity P9 simulation, panel D is the circular P9 simulation. Our area of interest (75\textless $a$\textless 200 au and $i$\textless 20\degree) is marked with the red boxes.   \label{fig:scatter}}
\end{figure*}

\begin{figure*}
\includegraphics[width=\textwidth]{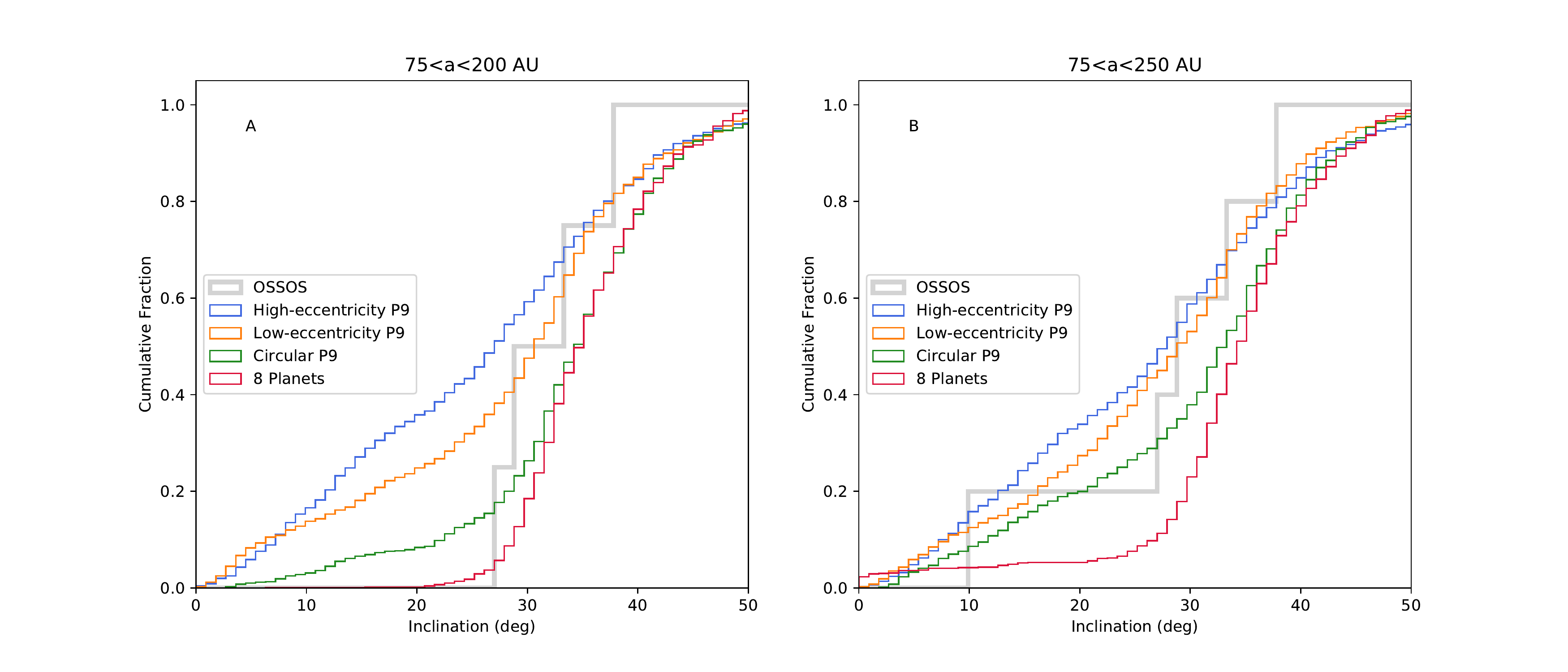}
%%\plotone{hist75.pdf}
\caption{Cumulative inclination distributions of detected objects with 40 \textless $q$ \textless 50 au. Panel A shows objects with semimajor axes between 75 and 200 AU, while Panel B shows objects with semimajor axes between 75 and 250 AU. Actual OSSOS detections are marked with the thick gray line, and synthetic detections are shown for 8 planets (red), high-eccentricity P9 (blue), low-eccentricity P9 (orange), and circular P9 (green).  \label{fig:hist}}
\end{figure*} 

In Figure \ref{fig:scatter}, we plot the inclinations of simulated detections of detached, distant objects for each simulation. To compare to the OSSOS-detected objects in orange, we co-add the outputs of each Myr over the last Gyr of each simulation and run this through the survey simulator to get 1000 simulated detections for each simulation. Shown here is the population generated with the contrast $H\textsubscript{r}$ distribution. The feature to highlight in Figure \ref{fig:scatter} is the near absence of objects with less than 20\degree inclination and semimajor axis below 200 au, marked with the red box in the figure. This region is also not occupied by any of the 5 OSSOS objects. However, the 9-planet simulations all have objects occupying this region. We ran a Kolmogorov-Smirnov (K-S) test on the inclination distributions of the outputs of the survey simulator for the range of semimajor axis of Figure \ref{fig:scatter} (75\textless $a$\textless 250 au) and the output for the range of 75\textless $a$\textless 200 au for the 8-planet and each of the 9-planet simulations and between the three 9-planet simulations. This was done for both the contrast and knee $H\textsubscript{r}$ distributions.  Each returned a result to allow us to reject the null hypothesis that they came from the same distribution at beyond the 3$\sigma$ confidence level. The Anderson-Darling test returned the same result for each of the simulations except the contrast distribution between the high- and low-eccentricity 9-planet simulations which could only be rejected at the 2$\sigma$ confidence level.

The standard mechanism in which objects become detached at these large ($\sim$50-200 AU) semimajor axes is via Kozai-Lidov cycles operating within Neptunian mean motion resonances. When an object in a resonance with Neptune exceeds a critical value, it will undergo Kozai cycles \citep{1962AJ.....67..591K,1962P&SS....9..719L,2003EM&P...92...29G}. During these cycles the object's eccentricity and inclination oscillate exactly out of phase. At high inclination, the eccentricity gets small while the semimajor axis remains fixed. This increases the object's perihelion, weakening the object's coupling to Neptune. The object can leave its resonance with Neptune and get locked into an orbit with a high inclination, and large perihelion, and semimajor axis as Neptune continues to migrate away from the Sun.

To better illustrate the difference in inclination distributions, we plot the cumulative inclination distributions in Figure \ref{fig:hist}. We plot the inclination distribution for the objects with 75 \textless $a$ \textless 200 au in panel (A) and 75 \textless $a$ \textless 250 au in panel (B). We consider both of these ranges of semimajor axis due to the low-inclination population that appears between 200 and 250 AU in the 8-planet simulation. Both populations are run through the survey simulator with the contrast $H\textsubscript{r}$ distribution. It can be seen in panel (A) that the high-eccentricity P9 and low-eccentricity P9 simulations produce a much larger low-inclination population than the 8-planet distribution. The circular P9 simulation also produces more low-inclinations objects than the 8-planet simulation, but at a diminished level. In panel (B) the 9-planet simulations are more similar to each other and all produce a larger low-inclination population than the 8-planet simulation. Using the K-S and Anderson-Darling tests, we can reject the null hypothesis that the inclination distributions of the simulations are from the same distribution in both ranges of semimajor axis. However, the small number of OSSOS objects (4 in panel A and 5 in panel B) prevent us from confidently rejecting or strongly favoring any of our models, as the results of the K-S and Anderson-Darling tests between the inclination distributions of each simulation and the OSSOS objects cannot reject the null hypothesis.
%%Although the 8-planet simulation appears to provide a nice match to the observational datasets in both panels, the small number of OSSOS objects (4 in panel A and 5 in panel B) prevent us from confidently rejecting or strongly favoring any of our models.

\begin{figure*}
\includegraphics[width=\textwidth]{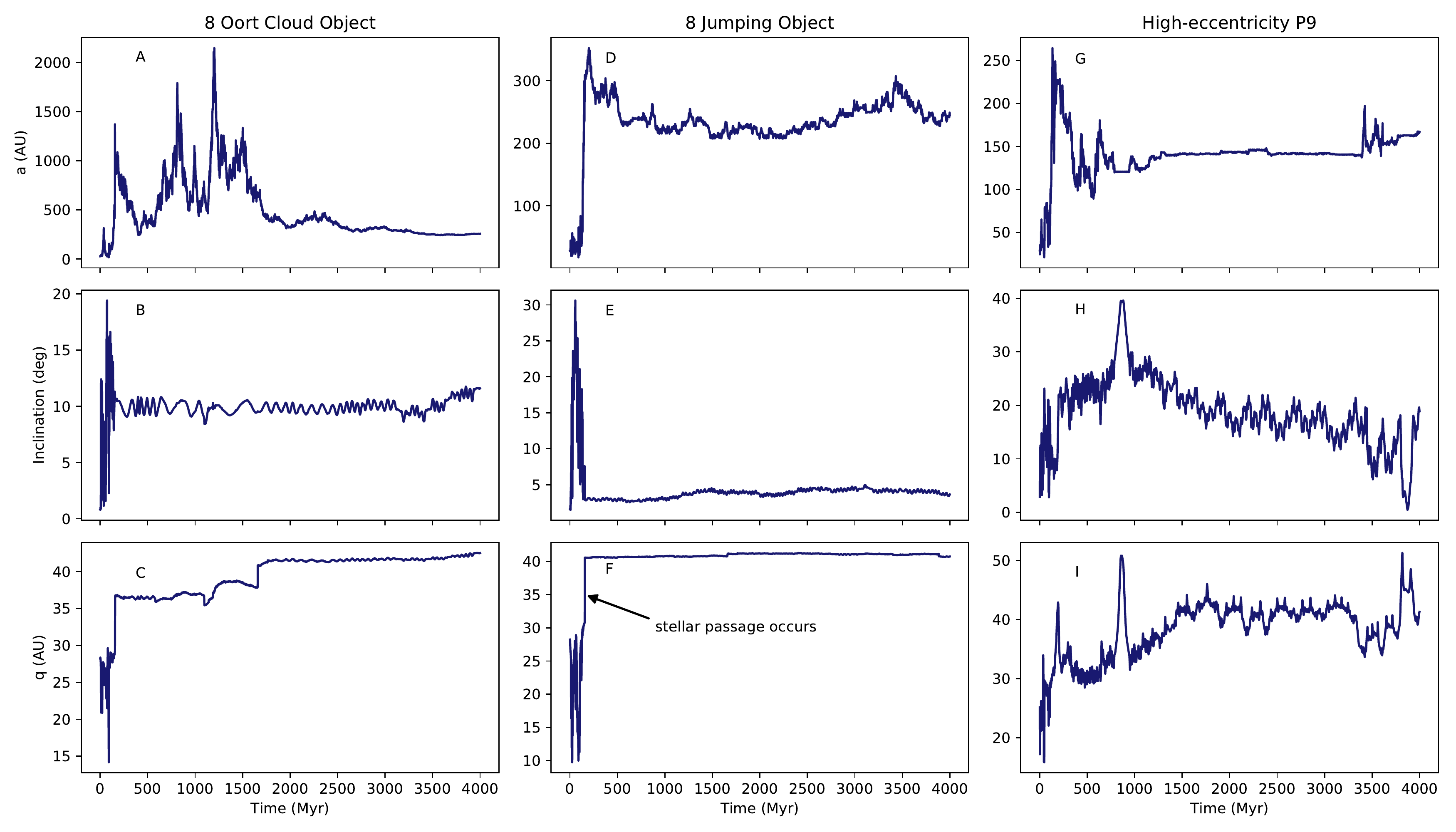}
%%\plotone{evolution.pdf}
\caption{Time evolution of three examples of objects that are in the low inclination, large $q$ (40-50 AU), and large semimajor axis (75-250 AU) orbits at the end of their respective simulations. The first column is the evolution of an object that moves out of the Oort Cloud back into the region of interest from the 8-planet simulation. The second column is the evolution of an object from the 8-planet simulation that has a jump in perihelion due to a close encounter from a passing star. The third column shows an object with an evolution driven by interactions with a 9th planet from the High-eccentricity P9 simulation. The first row shows semimajor axis, the second row shows inclination, and the third row shows perihelion.  \label{fig:evolution}}
\end{figure*}

In order to determine how many TNOs would need to be discovered in this distant, detached region of orbital space, we look at how many objects are needed in order to differentiate the different simulations. We run the distant, detached objects from each simulation through the survey simulator with the contrast $H\textsubscript{r}$ 10 times and then perform a K-S test on the inclination distributions of the 8-planet and each of the 9-planet simulations for an increasing number of detections for each simulation. We do this until we can reject the null hypothesis at the 2 and 3$\sigma$ confidence level. We average the number of detections between the 10 survey simulator results.
For objects in the range of semimajor axis 75\textless a \textless 200 AU and perihelion 40\textless q\textless 50, an average of 13 detections are needed to differentiate at the 2$\sigma$ confidence level the 8-planet simulation from the high-eccentricity P9 simulation, 27 for the low-eccentricity P9 simulation, and 201 for the circular P9 simulation. For a 3$\sigma$ confidence 23 are needed for the high-eccentricity P9, 47 for the low-eccentricity P9, and 349 for the circular P9. If we expand the range of semimajor axis to 75\textless a \textless 250 AU, we need 27 detections to differentiate the high-eccentricity P9 at the 2$\sigma$ level, 23 for the low-eccentricity P9, and 68 for the circular P9. For a 3$\sigma$ confidence level for this range of objects we need 40 detections for high-eccentricity P9, 46 for low-eccentricity P9, and 114 for circular P9 to differentiate from the 8-planet model. The Anderson-Darling test returns very similar numbers of required detections for each set of simulations.

In Figures \ref{fig:scatter} and \ref{fig:hist}(B), we also notice that every simulation possesses a significant population of objects with low inclination and high semimajor axis. These objects cannot be deposited in these orbits by the effects of Neptune and Kozai oscillations. There are two ways objects can attain these orbits without the effects of a ninth planet. We illustrate these two types of evolution in Figure \ref{fig:evolution}, with representative examples from our 8-planet simulation. The more common method in the 8 planet simulation is that the object gets trapped into the Oort Cloud with high semimajor axis and then has its semimajor axis slowly drawn down via weak planetary energy kicks into our region of interest. This type of evolution is illustrated in the leftmost column of Figure \ref{fig:evolution}.

The next evolution, which is displayed in the middle column of Figure \ref{fig:evolution} , is a less likely occurrence and a result of the closest stellar passage suffered by the solar system. \citet{2011Icar..215..491K} found that the most powerful stellar encounter(s) in the solar system's history can detach the perihelia of orbits interior to the traditional Oort Cloud, and this is what occurs for a handful of objects in our 8-planet run. This is the case for three such objects in our 8-planet simulation. These objects have a jump in perihelion from $\sim$30 to 40 au 156 Myr into the simulation. Their semimajor axes also increase to \textgreater 300 au. Weak-planet interactions then bring them back into the $\sim$ 200 au range with low inclination. The jump in perihelion is due to a rare powerful encounter with a passing star 156 Myr into the simulation. The star has a mass of 0.22 $M$\textsubscript{\(\odot\)}, velocity of 36 km s$^-1$, and an impact parameter of 540 AU. This is the closest stellar encounter of the entire simulation. Such an encounter's effectiveness at detaching the perihelia of TNOs depends on the encounter's timing and whether there are still a large number of scattering TNOs available to be detached. In our 8-planet simulation, this close encounter occurs very early in the simulation while Neptune is still migrating through the primordial Kuiper belt when there are still large numbers of scattering objects to be influenced by such an encounter. Although such a stellar encounter is expected over 4 Gyr, there is only a 7.5\% chance of it occurring while Neptune is still migrating and scattering large numbers of bodies. If it were to occur at a much later epoch, there would be far fewer actively scattering bodies for this encounter to detach from the region of the known planets. For most stellar encounter histories, we could therefore expect that low-inclination, detached orbits would be even more sparsely populated than the level seen in this particular 8-planet simulation.  

\begin{deluxetable*}{lccc}
\tablenum{2}
\tablecaption{Ratio of $i$ \textless 10\degree to $i$ \textgreater 25\degree \label{tab:ratio}}
\tablewidth{0pt}
\tablehead{
\colhead{Simulation} & \colhead{Pre-survey Simulator} & \colhead{Post-survey Simulator} &
\colhead{Post-survey Simulator} \\
\colhead{} & \colhead{} & \colhead{Contrast} & \colhead{Knee} 
\\
\colhead{(1)} & \colhead{2)} & \colhead{(3)} & \colhead{(4)} 
}
%\decimalcolnumbers
\startdata
8 planets & 0.00228 (0.0132) & 0.00102 (0.0454) & 0.00409 (0.0329) \\
High-eccentricity P9 & 0.105 (0.0984) & 0.270 (0.235) & 0.265 (0.265) \\
Low-eccentricity P9 & 0.0612 (0.0857) & 0.190 (0.183) & 0.130 (0.202) \\
Circular P9 & 0.0138 (0.0521) & 0.0323 (0.104) & 0.0357 (0.113) \\
\enddata
\tablecomments{Ratios of $i$ \textless 10\degree to $i$ \textgreater 25\degree for each simulation for the 75 \textless $a$ \textless 200 au semimajor axis range with the 75 \textless $a$ \textless 250 au case in parentheses. All ratios are for orbits with perihelia between 40 and 50 AU. Row 2 is the result without having been run through the survey simulator. The simulations in row 3 have been run though the survey simulator with the contrast $H\textsubscript{r}$ distribution. Row 4 simulations also have been run through the survey simulator with the knee $H\textsubscript{r}$ distribution.}
\end{deluxetable*}

Low-inclination objects within our range of semimajor axes are of course more likely to become detached when a ninth planet is present. In the rightmost column of Figure \ref{fig:evolution}, we show a typical example of this evolution. However, our particles do exhibit a wide variety of behavior. The dynamics of such bodies are complex, and both resonant and secular perturbations (in addition to Neptunian close encounters) play a role in populating our region of interest \citep{2016AJ....151...22B,2017AJ....154..229B,2018AJ....156..263L,2021arXiv210501065C}.

To further quantify the differences in the inclination distribution between the 8-planet and 9-planet simulations, we give the ratios of $i$ \textless 10\degree to $i$ \textgreater 25\degree for each simulation for both the 75 \textless $a$ \textless 200 au and 75 \textless $a$ \textless 250 au cases in Table \ref{tab:ratio}. One can see that even in the most conservative case the 9 planet ratios are a factor of $\sim$10 greater than the 8-planet ratio. There are only 4-5 objects in this range of semimajor axis now, but it is clear that the ratio of low inclination to high inclination in this range provides a strong lever on the types of undetected planets that could exist in our distant solar system. In the next several years, the Vera C. Rubin Observatory should be able to provide an ample number of further detections to confidently infer our solar system's true ratio and tightly constrain allowable 9-planet models \citep{2019ApJ...873..111I}.

\section{Conclusion} \label{sec:con}

Neptune's interactions with TNOs during and after its migration do not populate the low-inclination region of distant, detached TNOs. Because of this, there is a significant difference in the number of objects in this region between an 8-planet solar system and one that contains a distant 9th planet. Our 9-planet models populate this region considerably more than our 8-planet model. Even our most conservative 9-planet model (circular P9) produces a significant difference in the the number of low-inclination TNOs (relative to high-inclination ones) by a factor of ~5. The mechanisms that could place objects in this region without interactions with a 9th planet (migration in from the Oort Cloud and an unlikely early, very close encounter with a passing star) do not produce an abundance of low-inclination objects in this region while the secular perturbations from a 9th planet greatly increase the number of objects that end up in this region. 

These distant, detached objects can provide an important constraint on planet 9's properties as well as its existence. Generally, we expect the number of low-inclination bodies in this region of orbital space to increase with a distant planet's mass and eccentricity. 
%While the 5 OSSOS-detected objects currently known in this region bear a resemblance to the 8-planet model, m
More detected objects in this region of 75 \textless $a$ \textless 250 AU and 40 \textless $q$ \textless 50 AU are needed to adequately test these constraints on planet 9. Upcoming surveys, such as the Vera C. Rubin Observatory, should provide a large number of new detections in this region of interest and allow us to gain a better understanding of the outer reaches of our solar system.

\section{Acknowledgements} \label{sec:ack}

This work was supported under NSF award AST-1615975, NSF CAREER award 1846388, and NASA Emerging Worlds grant 80NSSC18K0600. The computing for this project was performed at the OU Supercomputing Center for Education \& Research (OSCER) at the University of Oklahoma (OU).

%% pdflatex paper.tex
%% bibtext paper
%% pdflatex paper.tex
%% pdflatex paper.tex

\bibliography{paper}
\bibliographystyle{aasjournal}

\end{document}